\documentclass[floatfix,aps,pra,showpacs,twocolumn]{revtex4}
\usepackage[T1]{fontenc}             
\usepackage{mathptmx}                
\usepackage{bm}
\usepackage{amssymb}
\usepackage{graphicx}
\usepackage{amsmath}
\usepackage{units}

\begin{document}

\title{Multiphoton ionization and stabilization of helium in superintense xuv fields}

\author{S. A. S{\o}rng{\aa}rd}
\affiliation{Department of Physics and Technology, University
  of Bergen, N-5007 Bergen, Norway}

\author{S. Askeland}
\affiliation{Department of Physics and Technology, University
  of Bergen, N-5007 Bergen, Norway}

\author{R. Nepstad}
  \affiliation{Department of Physics and Technology, University
  of Bergen, N-5007 Bergen, Norway}

\author{M. F{\o}rre}
\affiliation{Department of Physics and Technology, University
  of Bergen, N-5007 Bergen, Norway}

\date{\today}


\begin{abstract}

Multiphoton ionization of helium is investigated in the superintense field
regime, with particular emphasis on the role of the electron-electron
interaction in the ionization and stabilization dynamics.  To accomplish this,
we solve {\it ab initio} the time-dependent Schr{\"o}dinger equation with the
full  electron-electron interaction included.  By comparing the ionization
yields obtained from the full calculations with corresponding results of an
independent-electron model, we come to the somewhat counterintuitive conclusion
that the single-particle picture breaks down at superstrong field strengths.
We explain this finding from the perspective of the so-called
Kramers-Henneberger frame, the reference frame of a free (classical) electron
moving in the field.  The breakdown is tied to the fact that {\it shake-up} and
{\it shake-off} processes cannot be properly accounted for in commonly used
independent-electron models.  In addition, we see evidence of a change from the
multiphoton to the shake-off ionization regime in the energy distributions of
the electrons. From the angular distribution it is apparent that correlation is
an important factor even in this regime.

\end{abstract}

\maketitle

\section{Introduction} 
More than 20 years ago, theoretical studies of atomic hydrogen in
ultraintense, high-frequency laser fields produced an unexpected
result~\cite{Pont1988, Fedorov1988,
Kulander,Piraux1991, Gavrila1990, Su1990, Eberly1993, Geltman1994,Geltman1995}: when increasing the
intensity of the laser pulse to such a degree that the applied forces
dominate over the Coulomb attraction between the nucleus and the electron, the
ionization probability does not increase accordingly, but rather stabilize, or
start subsiding. This counterintuitive phenomenon was dubbed atomic
stabilization, and was subject to much research in the following decade.  The
discussions, controversies and conclusions are available in a number of
review articles, see e.g.~\cite{Gavrila2002, Popov2003, Fedorov2006} and
references therein.  It has also been pointed out that atomic stabilization has
a classical counterpart~\cite{Grochmalicki,Grobe1991}. (See
also~\cite{Gavrila2002} and references therein.)

At the start of the nineties, the laser technology required to experimentally
observe the stabilization effect in tightly bound systems was not available.
For example, in order to measure stabilization in atomic hydrogen, photon
energies exceeding $13.6$ eV, the binding energy of the atom, and intensities
on the order of $10^{16}$ W/cm$^2$ or more are
required~\cite{Dondera,Forre2005}.  Grobe and Eberly~\cite{Grobe1993}
demonstrated that stabilization could occur in H$^-$ at moderate intensities
($\sim10^{13}$ W/cm$^2$) and photon energies ($\sim 2$ eV), and Wei \textit{et
al.}~\cite{Wei2007} suggested an experiment, in which a laser, of realistic
frequency and intensity, could possibly stabilize the unstable He$^-$ ion.
However, at present, the only experimental confirmations of stabilization are
from studies of low-lying Rydberg
states~\cite{Boer1993,Druten1997,Talebpour1996,Hoogenraad1994}.  With recent
advances in free-electron laser (FEL) technology, extremely high peak
intensities have been achieved, with wavelengths ranging from vacuum
ultraviolet to soft x-rays~\cite{Ackermann,Shintake2008}, and even higher
intensities are expected to be delivered in the near future~\cite{Callegari}.
Thus, laser technology is approaching the regime needed for observing atomic
stabilization in ground state (neutral) atomic systems.

Although atomic stabilization has been studied extensively during the last two
decades, studies of stabilization in systems containing two electrons are still
scarce~\cite{Gavrila1996,Gavrila2002} and most often assessed with simplified
physical models of reduced
dimensionality~\cite{Grobe1993,Bauer1999,Staudt2003,Staudt2006}.  Including a
second electron adds a new dimension to the problem, manifested through the
electron-electron repulsion.  The studies mentioned above revealed that the
electron-electron interaction suppresses atomic stabilization in two-electron
systems. Although \textit{ab initio} calculations of helium have previously
been performed at fairly high intensities in the xuv
regime~\cite{Parker2001,Laulan2003a}, only recently was such endeavors extended
into the stabilization regime~\cite{Birkeland2010}, confirming the detrimental
effect of the electron-electron interaction on stabilization. However, it was
shown that the effect is markedly less than predicted in models of reduced
dimensionality.

In this work, we revisit the problem of multiphoton ionization of helium in
superintense high-frequency fields. In continuation of the work of Birkeland
{\it et al.}~\cite{Birkeland2010} we look more closely into the strong-field
ionization dynamics of the atom, with particular emphasis on atomic
stabilization, considering laser pulses of various central frequencies and
durations.  A comparison of the ionization yields obtained from the \textit{ab
initio} calculations, including correlation, with corresponding results
obtained from an independent-electron model reveals that the validity of the
latter breaks down at strong fields.  An analysis of the system equations in
the so-called Kramers-Henneberger
frame~\cite{Pauli1938,Kramers1956,Henneberger1968,Faisal} shows that the
electron-electron interaction plays a decisive role in this limit.  We further
show that this is manifested in the energy and angular distributions of the
ejected electrons.

Atomic units, where $m_e$, $\hbar$, and $e$ are scaled to unity, are used throughout unless stated otherwise.

\section{Methods}
    \subsection{\textit{Ab initio} calculations}
	We obtain the ionization probability of ground state helium in extreme laser fields from first principles, i.e.,
	by solving (numerically)
	the full time-dependent Schr\"odinger equation (TDSE). 
	Formulating the problem in the velocity gauge, the Hamiltonian assumes the form	
	\begin{equation}
		H = \sum_{i=1}^2 \left( \frac{\mathbf{p}_i^2}{2} - \frac{2}{r_i} + A_z(t)p_{z_i} \right)+\frac{1}{|\mathbf{r}_1-\mathbf{r}_2|}.
		\label{eq1}
	\end{equation}
	A sine-squared carrier envelope was chosen for the laser interaction, 
	\begin{equation}
		A_z(t)=A_0\sin^2\left(\frac{\pi t}{T}\right)\cos(\omega t).
		\label{}
	\end{equation}
	Here $A_0=E_0/\omega$, $E_0$ is the amplitude of the electric field, 
	$\omega$ is the laser frequency, and $T$ is the total pulse duration.
	The semiclassical treatment of the field is a valid approach due to the
	enormous photon flux of superintense lasers.  
	The pulse fulfills the constraint of a physical pulse~\cite{Muller1996},
    \begin{equation}
	    \int_0^T \mathbf{E}(t)dt=0.
	    \label{}
    \end{equation}
	Propagation and analysis of the wavefunction is performed with the PyProp
	framework~\cite{pyprop}, a Python/C++ software package for solving the TDSE.
	
	The wavefunction is expanded in a $B$-spline basis~\cite{Boor2001,
	Bachau2001} for each of the radial components, and a coupled
	spherical harmonic basis for the angular components,
	\begin{equation}
		\Psi (\mathbf{r}_1,\mathbf{r}_2,t)=\sum_{i,j,k}c_{ijk} \frac{B_i(r_1)}{r_1} \frac{B_j(r_2)}{r_2}
		\mathcal{Y}_{l_1,l_2}^{LM}(\Omega_1,\Omega_2),
		\label{eq:wavefunction}
	\end{equation}
	where $k = \{L,M,l_1,l_2\}$ is a combined index for the angular indices.
	The coupled spherical harmonic basis functions,
	\begin{equation}
		\mathcal{Y}_{l_1,l_2}^{LM}(\Omega_1,\Omega_2) = \sum_m \langle l_1l_2mM-m|LM\rangle Y_{l_1}^m(\Omega_1)Y_{l_2}^{M-m}(\Omega_2)
		\label{}
	\end{equation}
	are obtained  by linearly combining products of ordinary spherical harmonics, weighted by Clebsch-Gordan coefficients~\cite{Bransden2003}. 

	As the $B$-spline basis functions are not orthogonal, an overlap matrix
	$S_{ij}=\int B_i(r)B_j(r)dr$ is introduced for
	each electronic coordinate. From these the total overlap matrix is found
	for every angular momentum component by taking the Kronecker product
	$\mathbf{S}=\mathbf{I}_k\otimes\mathbf{S}_1\otimes\mathbf{S}_2$, where
	$\mathbf{I}_k$ denotes the identity matrix and $k$ is the angular index.
	The resulting TDSE may then be written as
	\begin{equation}
		i\mathbf{S} \frac{\partial}{\partial t}\mathbf{c}(t)=\mathbf{H}(t)\mathbf{c}(t)
		\label{}
	\end{equation}
	on matrix form.
	
	We solve the TDSE using a scheme based on the first-order approximation to
	the matrix exponential
	\begin{equation}
		\exp(-i\Delta t \mathbf{S}^{-1}\mathbf{H}) = \mathbf{I} - i\Delta t\mathbf{S}^{-1}\mathbf{H} + O(\Delta t^2).
		\label{}
	\end{equation}
	A direct application of this formula is not desirable due to numerical
	instabilities. Instead, we combine one half-step forward in time, 
	\begin{equation}
		\mathbf{c}(t+\Delta t/2) =\left(\mathbf{I}-\frac{i\Delta t}{2}\mathbf{S}^{-1}\mathbf{H}\right)\mathbf{c}(t),
	\end{equation}
	with one half-step backward in time, 
	\begin{equation}
		\mathbf{c}(t+\Delta t/2)=\left(\mathbf{I}+\frac{i\Delta t}{2}\mathbf{S}^{-1}\mathbf{H}\right)\mathbf{c}(t+\Delta t),
	\end{equation}
	to obtain the unconditionally stable Cayley-Hamilton form of the time propagator,
	\begin{equation}
		\left(\mathbf{S}+\frac{i\Delta t}{2}\mathbf{H}\right)\mathbf{c}(t+\Delta t) = \left(\mathbf{S}-\frac{i\Delta t}{2}\mathbf{H}\right)\mathbf{c}(t).
		\label{}
	\end{equation}
	This linear system of equations is too large to be solved directly, hence
	we use an iterative method. Since the matrix $(\mathbf{S}+\frac{i\Delta
	t}{2}\mathbf{H})$ is not Hermitian, our choice is the generalized minimum
	residual method (GMRES), a Krylov subspace method which combines Arnoldi
	iterations with a least-squares problem in the projected
	space~\cite{Saad1986,Saad2003}. In the GMRES algorithm the error in the
	least-squares residuals is controlled by the dimension of the Krylov
	subspace, which can be increased until the desired precision is obtained.

	\subsection{Calculating ionization}	
	In this work we compute the ionization probability resolved in direction
	and energy.	We also do a series of smaller simulations, calculating only
	the total ionization probabilities.  Separating the single and double
	ionization is achieved by projection onto double continuum
	states. In order to obtain these continuum states exactly one needs to
	solve a scattering problem for the full two-particle system.  As this is
	computationally cumbersome, an approximation using single-particle states
	is adopted instead.  It can be described as follows. In the case of double
	ionization, when both electrons are far from the nucleus, a product of
	continuum  He$^+$ $(Z=2)$ states is used. For single ionization, when one
	electron is close to the nucleus and the other far away a product of bound
	He$^+$ and continuum H $(Z=1)$ is used~\cite{Feist2008}.

	The single-electron states are not orthogonal to the
	bound states of the two-electron system, which may become populated
	during the action of the pulse. Therefore, the projection of
	the final wavefunction on the doubly bound states are removed before
	further analysis is conducted. Moreover, as the electron-electron correlation
	is neglected in the double continuum states, the system must be propagated
	after the pulse for all quantities to converge~\cite{Madsen2007}. 
	
	On the other hand, when only calculating the total ionization, a small
	radial box is sufficient.  It is no longer necessary to propagate the
	system after the pulse, in order to minimize the interaction term, nor to
	project onto continuum He$^+$ states.  An absorbing potential is applied at
	the box boundary in order to absorb the emitted electrons and to minimize
	reflection.  When coupled with an absorbing potential, we find that only
	about a third of the radius needed to resolve the differential
	probabilities is necessary. The total ionization probability is simply the
	complement of the probability of being in one of the bound states. 
	
	To find the bound states, we use the implicitly	restarted Arnoldi
	method~\cite{Sorensen1992}.  This is a version of the Arnoldi method for
	finding eigenpairs that refines the Krylov subspace basis in order to find
	the wanted eigenvectors and eigenvalues.  As the Arnoldi method tends to
	find the largest eigenvalues, we also use shifted inverse iterations, which
	let us find the eigenvalues near a given value.  

	Further details on the discretization, the time integration, and the analysis were presented in 
	a recent communication~\cite{Nepstad2010}.  

    \subsection{Independent-electron model}
	In order to gauge the importance of the electron-electron interaction, we repeat the
	calculations using an independent-electron (IE) model~\cite{Geltman1985}. The total wavefunction is approximated as a
	product of two single-electron wavefunctions
	\begin{equation}
	\label{eq2_2}
		\Psi(\textbf{r}_1, \textbf{r}_2) = \psi_{\textrm{SAE}}(\textbf{r}_1)\psi_{\textrm{He}^+}(\textbf{r}_2).
	\end{equation}
	The subscript SAE refers  to the single-active electron approximation. This is a common
	approximation for many-electron problems, which focuses on one electron at the time.
	Any dependence on the rest of the electrons is included in a common potential that is constant with regards to
	the other electron positions.  
	To find the first electron wavefunction $\psi_{\textrm{SAE}}$, we apply a pseudo potential, 
	which includes the shielding of the nucleus caused by the other
	electron~\cite{Tong2005}, 
	\begin{equation}
	    V(r) = - \frac{ Z + a_1 e^{-a_2r}+ a_3 r e^{-a_4r}+ a_5 e^{-a_6r} }{r}.
	    \label{eq:SAEpotential}
	\end{equation}
	For helium the effective charge $Z=1$ and the 
	coefficients $a_1=1.231$, $a_2=0.662$, $a_3=1.325$, $a_4=1.236$, $a_5=0.231$, and 
	$a_6=0.480$ were adopted. 
	The other electron moves in a He$^+$ potential, and it is therefore an
	accurate model for the singly ionized atom.	The IE model reproduces the
	correct ground state energies and single and double ionization thresholds,
	and decently represents the excited states.  As the name of the model
	suggests, the electrons do not interact with each other, beyond what is
	included in the shielded nuclear potential. That makes this a three-dimensional, 
	rather than a six-dimensional problem, and it can
	be calculated with relative ease on an ordinary computer.  As a consequence
	of working with independent particles, the total (single $+$ double) ionization probability becomes 
	\begin{equation}
		P_{\textrm{total}}^{\textrm{ion}} = 1 -  P^{\textrm{b}}_{\textrm{SAE}}P_{\textrm{He}^+}^{\textrm{b}}
	    \label{eq:SAEprobability}
	\end{equation}
	where $P^{\textrm{b}}_{\textrm{SAE}}$ and $P_{\textrm{He}^+}^{\textrm{b}}$ are the probability of the SAE and the He$^+$
	electron, respectively, being in a bound state. The probability for double ionization is obtained from the product 
	\begin{equation}
		P_{\textrm{double}}^{\textrm{ion}} = P^{\textrm{ion}}_{\textrm{SAE}}P_{\textrm{He}^+}^{\textrm{ion}}
	    \label{eq:DoubleProbability}
	\end{equation}
	and the single ionization probability is
	\begin{equation}
		P_{\textrm{single}}^{\textrm{ion}} = P^{\textrm{b}}_{\textrm{SAE}} P_{\textrm{He}^+}^{\textrm{ion}} 
						    + P_{\textrm{SAE}}^{\textrm{ion}}P^{\textrm{b}}_{\textrm{He}^+},
	    \label{eq:SingleProbability}
	\end{equation}
	where $ P^{\textrm{ion}}_{\textrm{SAE}}$ and $P_{\textrm{He}^+}^{\textrm{ion}}$ are the 
	ionization probabilities of the SAE and the He$^+$ electrons.

    \subsection{Convergence of the calculations}
	When doing the largest calculations, the radial domain typically extends to
	$\unit[80]{a.u.}$, although the double of this was employed to test the
	convergence. A 5th order  \textit{B}-spline basis of 185 splines is used,
	distributed exponentially near the nucleus, and linearly further away. Up
	to 300 splines were used for convergence test purposes.
	Regarding the angular basis of coupled spherical harmonics, $l \leq 5$ and
	$L \leq 6$ was found to be sufficient. Note that the system retains cylindrical
	symmetry in the presence of the $z$ polarized laser field.  Therefore, the $M$
	quantum number is set to 0 throughout.  Based on the calculations with a
	larger basis, the error is estimated to be less than 1\% in the ionization
	probabilities.
	
	For the smaller calculations, intended to provide only the total ionization probability, we
	use a small radial box of $\unit[30]{a.u.}$ and 80-100 \textit{B} splines of order 7, distributed
	linearly. Note that we have only a third the box size, but half the number of \textit{B} splines.
	In these calculations the angular basis went up to $l= 7$ and $L = 6$. The small box made it possible to go
	to higher intensities and pulse lengths than for the large box.
	The error in the ionization probability is gauged to be less than 5\% when
	$E_0 / \omega^2 > \unit[1]{a.u.}$, and less than 2\% for lower intensities.

\section{Results}

\subsection{Total ionization probabilities}
    \begin{figure}[t]
	    \begin{center}
		    \includegraphics[width=1.\columnwidth]{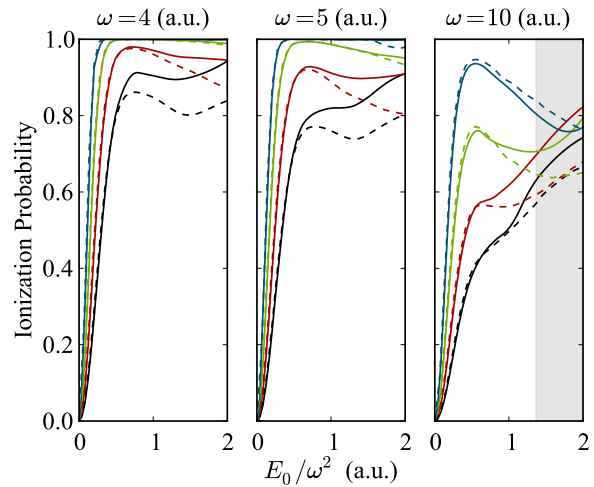}
		    \caption{(Color online) Ionization probabilities plotted as 
		    functions of the electronic displacement ($E_0/\omega^2$) for the
	    frequencies $\omega=\unit[4]{a.u.}$ (left), $\omega=\unit[5]{a.u.}$ (middle) and 
		    $\omega=\unit[10]{a.u.}$ (right). In each panel the pulses are
		    of 3, 6, 12, and 24 cycles duration from bottom to top. Dashed lines: IE calculations.} \label{fig1}
	    \end{center}
    \end{figure}

Figure~\ref{fig1} shows the total (single $+$ double) ionization probability
versus $\alpha_0$ for three different laser frequencies, $\omega=4$ (left
panel), $5$ (intermediate panel) and $10$ a.u. (right panel), and for four
different pulse durations, $3$, $6$, $12$ and $24$ cycles (from bottom to top).
Notice that on the abscissae, the domains are given in $\alpha_0 = E_0 /
\omega^2$, instead of intensity or peak electric field strength. Here
$\alpha_0$ represents the displacement amplitude of a free classical electron
in the oscillating field~\cite{Eberly1993}. This scaling allows us to easily
compare the results obtained with different laser frequencies.  In most of the
considered cases, the ionization probability increases with $\alpha_0$ up to
some point, where it attains a maximum before it starts to decline, i.e., we
are entering the so-called stabilization regime.  When stabilization occurs,
the ionization peak (corresponding to the 'death valley'~\cite{Gavrila2002}) is
typically situated between $\alpha_0 = 0.6$ and $\unit[0.7]{a.u.}$, independent
of laser frequency and pulse duration.  For very short pulse durations, e.g.
the 3-cycle pulse of $\omega = \unit[10]{a.u.}$, we observe a knee in the
function, rather than a peak at the stabilization point.  This is probably due
to the relatively large bandwidth of these short pulses and the averaging this
leads to.  For long pulse durations, e.g. the 24-cycle pulse of $\omega =
\unit[4]{a.u.}$, the atom is almost fully ionized, and the stabilizing effect
turns out to be weak.  The dashed lines in the figure are the results of the
independent-electron model.  They show good agreement with the full
calculations for weak fields ($\alpha_0 < \unit[0.5]{a.u.}$), and for long
pulses, but otherwise tend to overestimate the stabilization.  As a matter of
fact, the results show that the electron-electron interaction suppresses
stabilization in all cases.  We will return to the reason for this later.     
    
    \begin{figure}[t]
	    \begin{center}
		    \includegraphics[width=1.\columnwidth]{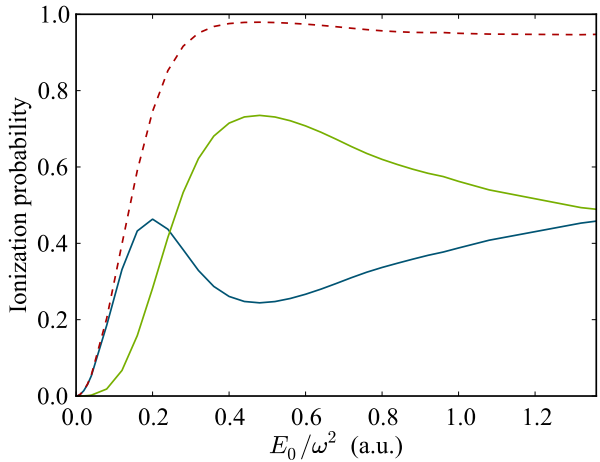}
		    \caption{(Color online) Ionization probabilities for a constant pulse duration of
		    $\unit[2\pi]{a.u.}$ The lines correspond to laser frequencies of 
		    $\omega~=~4,6,8,\unit[10]{a.u.} $, from top to bottom, or equivalently pulse lengths of 4, 6, 8, 10
		    cycles. 
		    The dashed lines are the IE calculations. 		 
		    } \label{fig2}
	    \end{center}
    \end{figure}

Figure~\ref{fig2} shows the ionization probability as a function of $\alpha_0$
for a pulse of constant duration $T =\unit[2\pi]{a.u.}$, but for varying
frequencies, $\omega=4$, $6$, $8$, and $10$ a.u. The corresponding results of
the IE model are shown in dashed lines.  One immediately perceives that for
higher frequencies, the atom stabilizes at lower ionization probabilities, in
accordance with the results in Fig.~\ref{fig1}.  Note that in the limit of weak
fields single ionization is by far the dominating ionization channel. Thus,
from first order perturbation theory $P_{\textrm{total}}^{\textrm{ion}} \propto
\alpha_0^2 T$.  Now, since the pulse duration is kept fixed in  Fig.~\ref{fig2}
(as opposed to Fig.~\ref{fig1}), this explains why the results of the
calculations with different frequencies almost coincide at smaller fields.  The
figure  also demonstrates the fact that the discrepancy between the IE model
(dashed lines) and the full calculations (solid lines) increases with the
intensity.  Furthermore, the stabilizing effect turns out to be very weak in
the fully correlated system.  Whereas the full \textit{ab initio} calculations
give ionization probabilities that level off (low frequencies) or increase
(high frequencies) for high intensities, the IE model returns probabilities
that are noticably lower.  As the intensity grows, so do the discrepancy.  As
such, the simplified model tends to always underestimate  the ionization
probability, with the consequence that the stabilization effect is
overestimated.

\subsection{The role of electronic correlation}
Figures~\ref{fig1} and~\ref{fig2} clearly demonstrate that the 
validity of the independent-electron
model~(\ref{eq2_2}) breaks down in the superintense field regime. This may appear
counterintuitive, as one might well expect the opposite to happen, i.e., that
the  importance of the electron-electron interaction should be negligible in the
presence of a strong external perturbation.
The reason why the electron-electron interaction in fact becomes more
important in this limit can be understood by analyzing the
dynamics in the so-called Kramers-Henneberger (KH)
frame~\cite{Pauli1938,Kramers1956,Henneberger1968,Faisal}, the rest frame of a
classical free electron in the laser field.  In this frame the
Hamiltonian, Eq.~(\ref{eq1}), is cast into the 
form,              
\begin{equation}
H_{\textrm{KH}} = \sum_{i=1}^2 \left( \frac{\mathbf{p}_i^2}{2} + V_{\textrm{KH}}\left[\mathbf{r}_i+ 
\bm{\alpha}(t)\right]\right)+
\frac{1}{\left |\mathbf{r}_1-\mathbf{r}_2\right|}, 
\label{eq2}
\end{equation}
where
\begin{equation}
\label{eq3}
V_{\textrm{KH}}\left[\mathbf{r}_i+ \bm{\alpha}(t)\right]=-\frac{2}{\left | \mathbf{r}_i+ 
\bm{\alpha}(t)\right |},
\end{equation}
is the Kramers-Henneberger potential, and
\begin{equation}
\bm{\alpha}(t)=\int_0^t A_z(t') dt' \hat{\bm{z}}
\label{eq4}
\end{equation}
represents the position relative to the laboratory frame 
of a classical free electron in the electric field $E_z(t)=-\partial A_z/\partial t$. 
One characteristic feature of the KH frame is that the dipole
interaction terms enter into the electron-nucleus Coulomb potentials
[c.f. Eq.~(\ref{eq3})], 
which in turn become time-dependent and modified by the external field.
Note also that the electron-electron interaction term is left
unaffected by the HK transformation. Assuming for the moment
that the Hamiltonian is periodic in time, i.e., neglecting the pulse profile,
the KH potentials, Eq.~(\ref{eq3}), are expanded in a Fourier series as
\begin{equation}
\label{eq5}
V_{\textrm{KH}}\left[\mathbf{r}_i+ \bm{\alpha}(t)\right]=
\sum_{n} V_n(\alpha_0,\mathbf{r}_i)e^{-in\omega t},
\end{equation}
with
\begin{equation}
	\label{eq6}
	V_n(\alpha_0,\mathbf{r}_i)   =
	\frac{1}{T}\int_0^{T}e^{-in\omega t}V_{\textrm{KH}}\left[\mathbf{r}_i+ \bm{\alpha}(t)\right]dt.
\end{equation}

Inserting the expansion~(\ref{eq5}) into the TDSE and applying high-frequency
Floquet theory, Gavrila {\it et
al.}~\cite{Gavrila1984,Gavrila_book,Gavrila2002,Gavrila2008} showed that the
$n=0$ component in Eq.~(\ref{eq6}) plays an increasingly important role in the
dynamics at higher value of $\alpha_0$. Furthermore, in the limit of
superintense fields ($\alpha_0\gg1$), F{\o}rre {\it et al.}~\cite{Forre2005}
showed that the ionization dynamics of atomic hydrogen is mainly dictated by
the $V_0$ potential.  Thus, in this limit, the dynamics of the system is
approximately given by the effective Hamiltonian 
\begin{equation}
	H_{\textrm{KH}}^{\textrm{eff}} = \sum_{i=1}^2 \left( \frac{\mathbf{p}_i^2}{2} +V_0(\alpha_0,\mathbf{r}_i)\right) +
	\frac{1}{\left |\mathbf{r}_1-\mathbf{r}_2\right|}.
	\label{eq7}
\end{equation}
Note that this Hamiltonian is time-independent and accounts
for {\it shake-up} (excitation) and {\it shake-off} (ionization)
in the strong field limit. 

An analysis of the properties of the $V_0$ potential term in the vicinity of
the origin reveals that it can be neglected relative to the electron-electron
repulsion term in the limit  $\alpha_0\rightarrow \infty $, provided the
two-electron wave function is localized, i.e., $\langle r_1\rangle\ll \alpha_0$
and $\langle r_2\rangle\ll \alpha_0$.
This means that the dynamics of the two-electron system, in the limit of
superintense fields and for sufficiently short pulses, ultimately reduces to
that of a pure Coulomb explosion process effectuated by the Coulomb repulsion
term in Eq.~(\ref{eq7}).  From this we conclude that the electron-electron
interaction is in fact very important in the strong field limit, effectively
reducing the stabilization effect.  Returning to the laboratory frame of
reference this should be understood in the following way: In the very strong
field limit, the electrons effectively behave like free particles in the field,
moving side by side with respect to the field axis. As this happens, the
nuclear attraction may become less important than the mutual repulsion between
the electrons, and the ionization is most likely initiated by electron-electron
scattering events (Coulomb explosion) and not electron-nucleus collisions.
This explains qualitatively why the ionization probabilities, calculated within
the independent-electron model, deviate increasingly from the exact ones in the
limit of stronger fields (c.f.  Figs.~\ref{fig1} and~\ref{fig2}).  As such, the
observed deviation is indeed a manifestation of the breakdown of the
single-particle picture in superstrong fields.

Notice that the effective Hamiltonian in Eq.~(\ref{eq7}) only depends indirectly
on the laser frequency through the displacement amplitude $\alpha_0$,
explaining why the validity of the IE model in Fig.~\ref{fig2}  breaks down at
approximately the same value of $\alpha_0$ independent of the laser frequency.

    \begin{figure}[t]
	    \begin{center}
		    \includegraphics[width=1.\columnwidth]{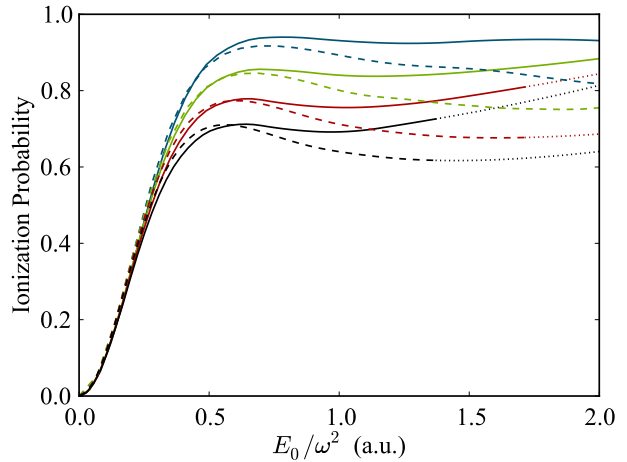}
		    \caption{(Color online) Excitation probabilities for the same scenario as in Fig. 
		    \ref{fig2}. The laser frequencies are $\omega = 10, 8, 6, \unit[4]{a.u.}$, from
		    top to bottom. The dashed lines are results from the IE model, while the solid lines are
		    results from the full calculations.
		    } \label{fig3}
	    \end{center}
    \end{figure}

Figure~\ref{fig3} shows the probability of excitation of helium for the cases
considered in Fig.~\ref{fig2}. Comparing Fig.~\ref{fig2} and~\ref{fig3} we
observe that the   decreasing ionization probability in the stabilization
regime is accompanied by a corresponding increase in the excitation
probability. Note that due to the high photon energy, excitation is here caused
by shake-up processes, merely demonstrating the importance of the $V_0$
potential in the stabilization regime.  The figure also clearly expresses the
fact that shake-up is more important for the higher frequencies and that the IE
model fails in describing shake-up (and shake-off)  processes accurately, in
accordance with the KH discussion above.

	\subsection{Analysis of angular and energy distributions}
	Further insight into the strong-field behavior of helium may be gained by
	examining the energy distribution of the ejected electrons. In particular,
	imprints left by the electron interaction in the angular distribution of the
	outgoing electrons may give further clues as to its importance at the
	different field strength regimes considered here.

    \begin{figure}[th]
	    \begin{center}
		    \includegraphics[width=.8\columnwidth]{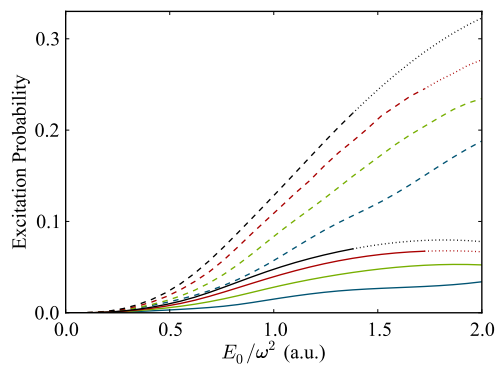}
			\caption{(Color online) Double ionization energy distribution for
			an $\omega = 5$ a.u. 6-cycle pulse ($\unit[182]{as}$), and for pulse
			amplitudes of 1 a.u. (upper panel), 10 a.u. (middle panel) and 20
			a.u. (lower panel). 
			}
			\label{fig:dpde_double_panels}
	    \end{center}
    \end{figure}
	%
	In Fig.~\ref{fig:dpde_double_panels} the double ionization energy
	distribution is shown for three different pulse amplitudes, 1, 10 and, 20
	a.u (from top to bottom).  The pulse duration was fixed at 6 cycles
	($\unit[182]{as}$), with a frequency of $\unit[5]{a.u.}$ At the lower
	intensity (amplitude), one-photon ionization dominates as expected.	 As the
	intensity increases, two-photon ionization becomes prominent, and
	higher-order double-electron above threshold ionization (DATI)	peaks start
	to appear~\cite{Parker2001}.  Since the one-photon process is highly
	correlated and depends critically on exchange of energy between the two
	electrons, it becomes less important at stronger fields, and two-photon
	double ionization takes over as the dominating channel.  At the highest
	intensity, more structures appear in the energy spectrum, caused by
	sidebands in the pulse, and the one-photon ionization process has become
	negligible.	

	The two-photon DATI component manifests itself as a single-peaked structure
	in Fig.~\ref{fig:dpde_double_panels}, in contrast to the common double-peak
	structure associated with sequential ionization~\cite{Parker2001}.  With
	the ultrashort pulse considered here, the second photon is absorbed before
	the residual ion has had time to relax to the ground state, but if the
	duration is increased to beyond $20$ cycles, relaxation may occur and a
	double-peak structure appears (not shown here).  The fact that the two
	peaks, corresponding to sequential two-photon double ionization in the long
	pulse limit, shift towards each other in the short pulse regime, and
	eventually merge into one single peak (located at equal energy sharing) is
	well known and has been studied in a series of papers in the weak  field
	(perturbative)
	limit~\cite{Laulan2003a,Laulan2003,Ishikawa2005,Barna2006,Foumouo2006,Foumouo2008,Palacios2009,Feist2009,Palacios2010,Foumouo2010,Pindzola2009}.
	The results in Fig.~\ref{fig:dpde_double_panels} demonstrate that this
	feature survives in the superintense field regime, representing a clear
	departure from the independent-electron model Eq.~(\ref{eq2_2}).

    \begin{figure}[ht]
	    \begin{center}
		    \includegraphics[width=1.\columnwidth]{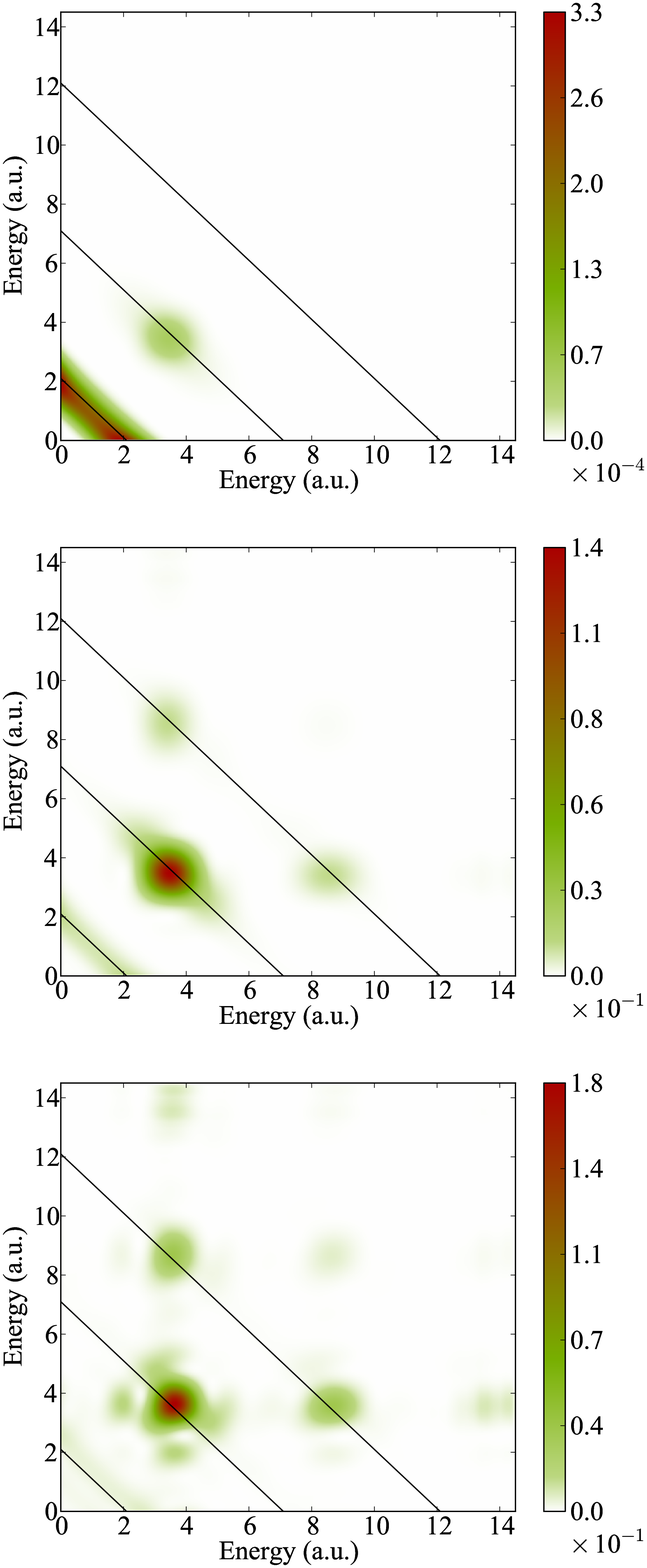}
		    \caption{(Color online) Angular distributions for double
			ionization with equal energy sharing $E_1=E_2 = (2\omega - I_p)/2$.
			The arrow indicates the fixed direction of the first electron
			. Solid (blue) line: $E_0 =
			\unit[1]{a.u.}$ Dashed (green) line: $E_0 =
			\unit[10]{a.u.}$ Dotted (red) line: $E_0 =
			\unit[20]{a.u.}$}
			\label{figX}
	    \end{center}
    \end{figure}

Figure~\ref{figX} shows the conditional angular distributions of the ejected
electrons obtained at the two-photon DATI peak  in
Fig.~\ref{fig:dpde_double_panels}, with equal energy sharing, and one of the
electrons emitted along the polarization direction (indicated with an arrow in
the figure).  The figure clearly shows that the distribution has a
backward-forward asymmetry even at the highest intensity considered,
demonstrating the breakdown of the single-particle picture, wherein a symmetric
double lobe (dipole) distribution would be found. The results are in accordance
with recent results obtained at weaker
fields~\cite{Feist2009,Palacios2009,Palacios2010}, and shows that the
back-to-back ejection mechanism is largely preserved even at very strong
fields. 

    \begin{figure}[th]
	    \begin{center}
		    \includegraphics[width=1.\columnwidth]{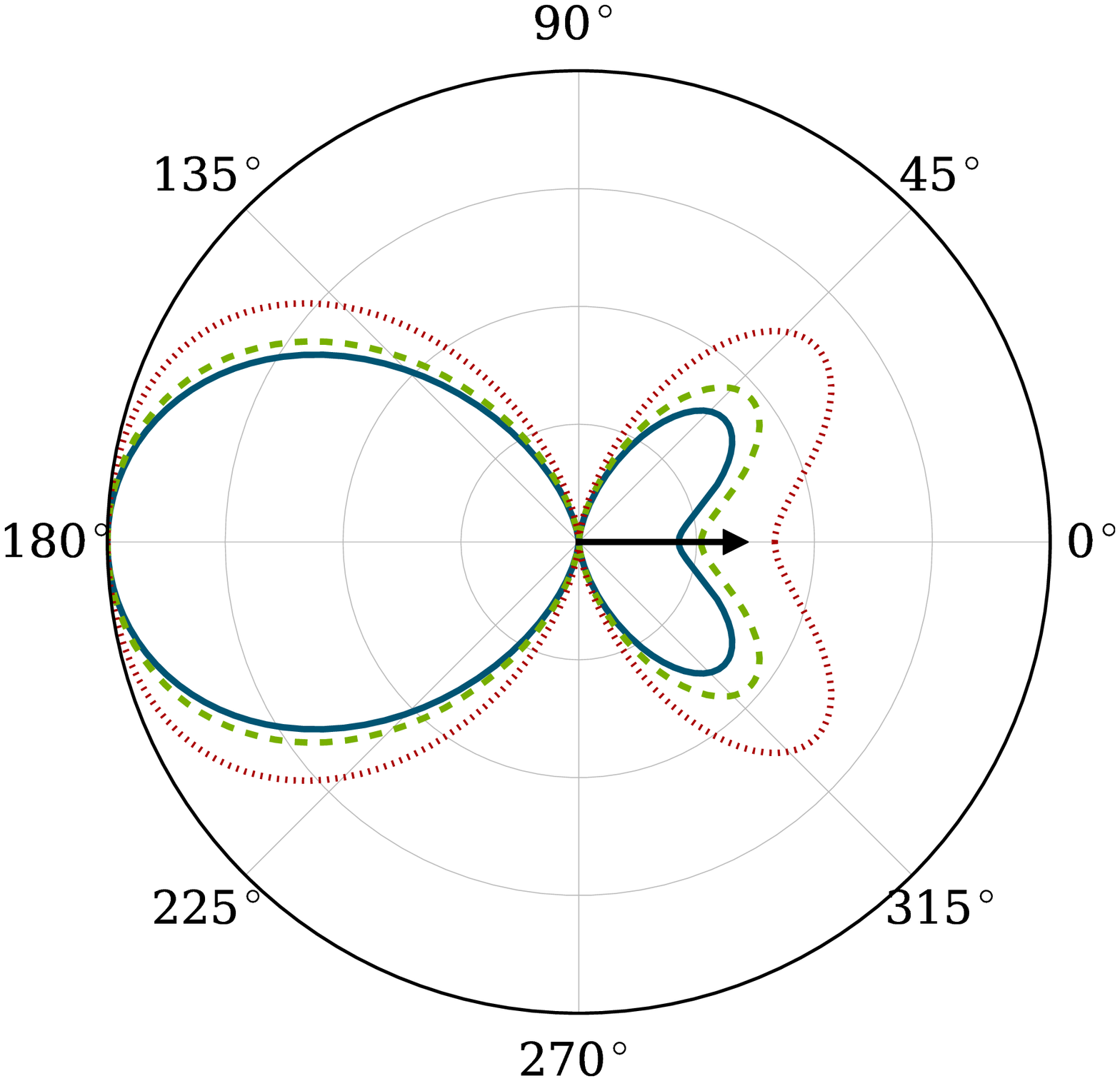}
		    \caption{(Color online) Energy distributions as a function of laser
				field strength. Top panel: He$^+$. Bottom panel: helium (single
				ionization). 
				See text for details.}
			\label{fig:dpde_intensity_dep}
	    \end{center}
    \end{figure}

	From numerical studies of stabilization in atomic hydrogen, it is known
	that the stabilization phenomenon is accompanied by the appearance of slow
	electrons~\cite{Forre2005, Toyota2009}. As the intensity is increased
	beyond the ionization maximum, where stabilization sets in, and for
	sufficiently short pulses, a peak structure near zero energy appears in the
	electron energy spectrum, becoming increasingly dominant as the intensity
	becomes large. This may be understood from the Kramers-Henneberger analysis
	above, and the importance of the $V_0$ potential in the limit of short,
	intense pulses. 
	In order to provide a baseline comparison for the two-electron case
	considered here, we have calculated the energy distribution for ionization
	of He$^+$, with identical pulse characteristics as those used in
	Fig.~\ref{fig:dpde_double_panels}. The result is shown in the upper panel of
	Fig.~\ref{fig:dpde_intensity_dep}.  We note the presence of ATI peaks, and,
	at the highest intensities, a slow electron peak (SEP) near zero
	energy~\cite{Forre2005, Toyota2009}. 
	The corresponding single ionization energy distribution of helium is shown
	in the lower panel of Fig.~\ref{fig:dpde_intensity_dep}, and indeed, a
	slow electron peak is visible. Note that the onset of slow electrons occurs
	at lower field strengths in the single ionization of helium than in He$^+$,
	which is related to the different ionization potentials ($I_p$). 

\begin{figure}[th]
	    \begin{center}
		    \includegraphics[width=1.\columnwidth]{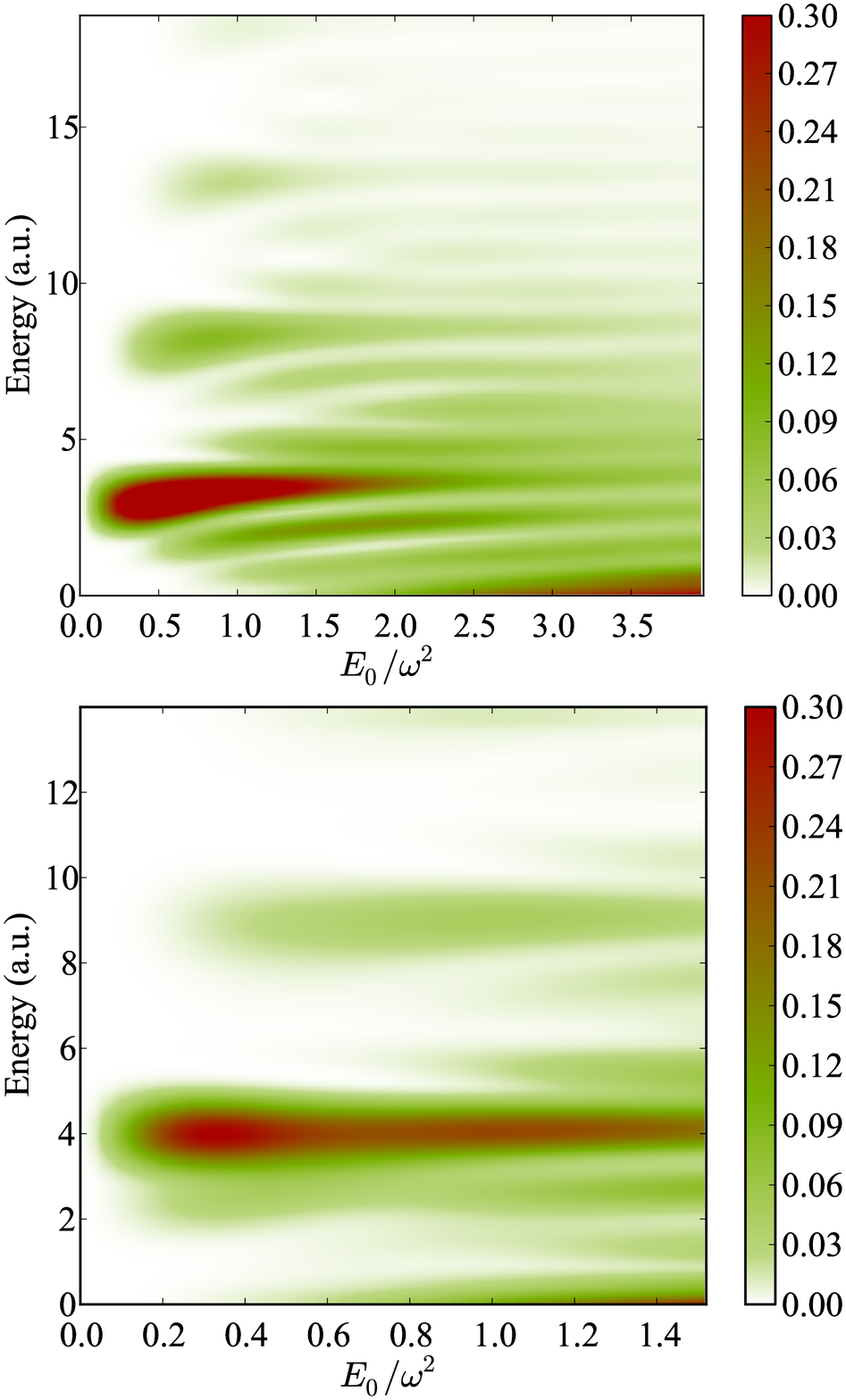}
		    \caption{(Color online) Ratio of slow to fast electrons for the
			double ionization process, shown for different pulse frequencies
			and durations, and plotted as a function of
			$\alpha_0$. The blue bar indicates the region where the
			corresponding double ionization probability is maximum; where
			stabilization sets in.}
			\label{fig6}
	    \end{center}
    \end{figure}

	Now, examining the lower panel in Fig.~\ref{fig:dpde_double_panels}, it
	appears that slow electrons do not emerge in the double ionization process
	at this intensity. However, the ionization potential is greater than that
	for single ionization, and therefore higher intensities are needed to reach
	the regime where a SEP may appear. Since He$^+$, with an ionization
	potential of $I_p = \unit[2]{a.u.}$, exhibits an onset of slow electrons around
	50 a.u., similar or possibly even higher field strength may be required for
	a SEP to appear in the double ionization of helium.  We may, however,
	observe the onset of slow electrons by partitioning the double ionization
	energy distribution into a low- to high-energy part, and considering the
	ratio of these two, c.f. Fig.~\ref{fig6}. In this figure, the ratio
	$P(E<E_c)/P(E>E_c), \ E_c = 3/2\omega - I_p$, for different frequencies and pulse
	durations are shown, and in all cases, we observe an increase of low-energy
	electrons after the stabilization peak (indicated by the blue bar),
	however it is most pronounced for the shorter pulses.

	\section{Conclusion}
	In conclusion, we have presented an in-depth analysis of two-electron
	dynamics driven by high-intensity ultrashort laser pulses in the xuv
	regime. Expanding on our earlier investigation of correlation effects in
	the stabilization of helium, we have shown that stabilization occurs within
	a narrow interval of values of $\alpha_0$, independent of frequency and
	pulse duration. This is also the point at which an independent-electron
	picture begins to break down, demonstrating the important role of the
	electron-electron interaction at high intensities. Through an analysis of a
	high-intensity limit form of the Hamiltonian, expressed in the
	Kramers-Henneberger frame, this feature may be understood. Further
	indications of intense field correlation effects are found in the angular
	distributions, where a backward-forward asymmetry is found for a wide range
	of intensities.  Finally, we have shown that slow electrons emerge at high
	intensities, as they do in one-electron systems, but at different
	intensities for single and double ionization.

	\section*{Acknowledgments}
	This work was supported by the Bergen Research Foundation (Norway). The
	calculations were performed on the Cray XT4 (Hexagon) supercomputer at
	Parallab, University of Bergen (Norway). CPU-hours were provided by NOTUR.


\end{document}